\documentclass{article}
\usepackage{spconf,amsmath,graphicx}

\usepackage{cite}
\usepackage[figurename=Fig.]{caption}
\usepackage{amsmath,amssymb,amsfonts}
\usepackage{graphicx}
\usepackage{textcomp}
\usepackage{xcolor}
\usepackage{balance}
\usepackage{cite}
\usepackage{hyperref}
\usepackage{amsmath,amssymb,amsfonts}
\usepackage{amsmath}
\usepackage{amsthm}
\DeclareMathOperator*{\argmax}{argmax}

\usepackage{graphicx}
\usepackage{textcomp}
\usepackage{xcolor}
\usepackage{graphicx}
\usepackage{float}
\usepackage{subfigure}
\usepackage{amsmath}
\usepackage{amsfonts,amssymb,yfonts}
\usepackage{mathrsfs}
\usepackage{mathtools}
\usepackage{bm}
\usepackage{multirow}
\usepackage{array}
\usepackage{amssymb}
\usepackage{amsmath}
\usepackage{cite}
\usepackage{url}
\usepackage{xcolor}
\usepackage{cite,graphicx,amsmath,amssymb}
\usepackage{subfigure}
\usepackage{fancyhdr}
\usepackage{mdwmath}
\usepackage{mdwtab}
\usepackage{caption}
\usepackage{amsthm}
\usepackage{setspace}
\usepackage{bm}
\usepackage{mathtools}
\usepackage{dsfont}
\usepackage{bbm}
\usepackage{algorithm}
\usepackage{algorithmic}

\newtheorem{theorem}{Theorem}

\newtheorem{lemma}{Lemma}

\newtheorem{corollary}{Corollary}

\DeclareMathAlphabet\mathbffrak{OMS}{cmsy}{b}{n}
\makeatletter
\newcommand{\biggg}{\bBigg@{3}}
\newcommand{\Biggg}{\bBigg@{3.5}}
\makeatother

\title{Enabling Secure Wireless Communications via Movable Antennas}
\name{Zhenqiao Cheng$^{\dag}$, Nanxi Li$^{\dag}$, Jianchi Zhu$^{\dag}$, Xiaoming She$^{\dag}$, Chongjun Ouyang$^{\ddag}$, and Peng Chen$^{\dag}$
\address{$^{\dag}$6G Research Centre, China Telecom Beijing Research Institute, Beijing, 102209, China\\
$^{\ddag}$University College Dublin, Dublin, D04 V1W8, Ireland}\thanks{This work was supported by the Young Elite Scientists Sponsorship Program by CAST under Grant 2022QNRC001.}}
\begin{document}
%
\maketitle
\begin{abstract}
A pioneering secure transmission scheme is proposed, which harnesses movable antennas (MAs) to optimize antenna positions for augmenting the physical layer security. Particularly, an MA-enabled secure wireless system is considered, where a multi-antenna transmitter communicates with a single-antenna receiver in the presence of an eavesdropper. The beamformer and antenna positions at the transmitter are jointly optimized under two criteria: power consumption minimization and secrecy rate maximization. For each scenario, a novel suboptimal algorithm was proposed to tackle the resulting nonconvex optimization problem, capitalizing on the approaches of alternating optimization and gradient descent. Numerical results demonstrate that the proposed MA systems significantly improve physical layer security compared to various benchmark schemes relying on conventional fixed-position antennas (FPAs).
\end{abstract}
\section{Introduction}
\label{sec:intro}

Multiple-input multiple-output (MIMO) technology plays a pivotal role in addressing the burgeoning data traffic demands within wireless communications \cite{Heath2018}. By harnessing multiple antennas, MIMO effectively exploits the spatial resources of wireless channels, resulting in substantial enhancements in spectral efficiency (SE). However, in conventional multiple-antenna systems, antennas remain fixed in position. This limitation restricts their ability to fully exploit the spatial variations within wireless channels within a given transmit/receive area, particularly when the number of antennas is limited \cite{Zhu2022}.

To harness additional spatial degrees of freedom (DoFs) and further enhance SE, the concept of movable antennas (MAs) has garnered increasing attention \cite{Zhu2022,Zhu2023,Ma2022,Zhu2023_2,Cheng2023_1,Cheng2023_2,Sun2023,Ma2023}. By connecting MAs to RF chains via flexible cables and allowing their positions to be adjusted in real-time through controllers, such as stepper motors or servos \cite{Ismail1991,Basbug2017}, MAs depart from the constraints of conventional fixed-position antennas (FPAs) \cite{Zhu2022}. This flexibility enables MAs to dynamically adapt their positions, thereby reshaping the wireless channel to achieve superior wireless transmission capabilities \cite{Zhu2022}.

Meanwhile, wireless networks confront a pressing concern stemming from the inherently open nature of the wireless medium: the formidable challenge of information leakage resulting from eavesdropping attacks \cite{Khisti2010}. This security threat has catalyzed the emergence of physical layer security mechanisms. Recently, the efficacy of multiple-antenna techniques and secure beamforming designs have been empirically validated as formidable tools for fortifying security \cite{Chen2017}. 

Within this context, it is a natural progression to amalgamate the realms of physical layer security and MAs to amplify the anti-eavesdropping capabilities of wireless communication systems, offering promising avenues for further research. However, the exploration of MA-enabled secure transmission remains relatively inchoate. As an initial attempt, we investigate the joint optimization of transmit MA positions and the transmit beamforming vector for improving the physical layer security of a multiple-input single-output single-antenna eavesdropper system, with MAs equipped at the transmitter. Our primary contributions are summarized as follows: {\romannumeral1}) We propose an MA-enabled secure transmission framework that harnesses the MAs to optimize antenna positions for bolstering the physical layer security. {\romannumeral2}) Under the criteria of secrecy rate maximization and power consumption minimization, we propose efficient iterative algorithms to jointly optimize the MA positions and transmit beamformer. {\romannumeral3}) Simulation results demonstrate that the proposed MA-based secure transmission provides more DoFs for improving physical layer security than conventional FPA-based ones.
\vspace{-10pt}
\section{System Model}
\vspace{-5pt}
\subsection{System Description}
\begin{figure}[!t]
\centering
\subfigbottomskip=0pt
	\subfigcapskip=-5pt
\setlength{\abovecaptionskip}{0pt}
\includegraphics[height=0.18\textwidth]{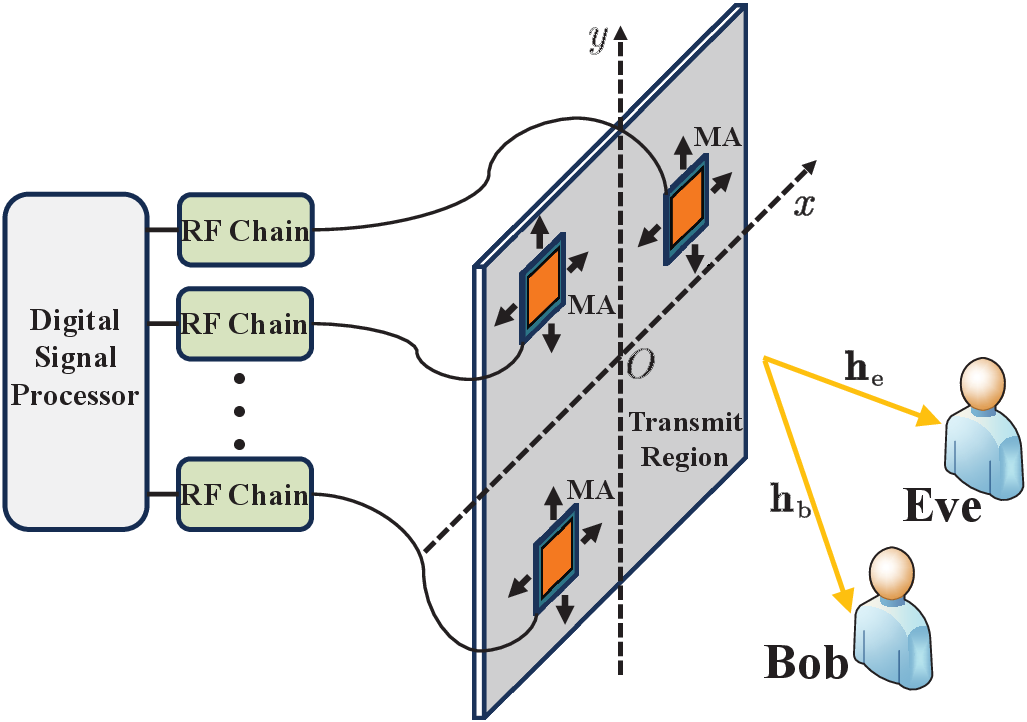}
\caption{The MA-enabled secure communication system}
\label{System_Model_MISOSE}
\vspace{-20pt}
\end{figure}

\vspace{-5pt}
We consider secure transmission in an MA-enabled setting as depicted in {\figurename} {\ref{System_Model_MISOSE}}, where a transmitter with $N$ MAs sends the securely coded message to a single-antenna legitimate user (Bob, denotes by $\rm{b}$) in the presence of a single-antenna eavesdropper (Eve, denoted by $\rm{e}$). The transmit MAs are connected to RF chains via flexible cables, and thus their positions can be adjusted in real time. The positions of the $n$th MA can be represented by Cartesian coordinates ${\mathbf{t}}_n=[x_n, y_n]^{\mathsf{T}}\in{\mathcal{C}}$ for $n=1,\ldots,N$, where $\mathcal{C}$ denotes the given two-dimensional region within which the MAs can move freely. Without loss of generality, we set $\mathcal{C}$ as square regions with size $A\times A$.

We assume quasi-static block-fading channels, and focus on one particular fading block with the multi-path channel components at any location in $\mathcal{C}$ given as fixed. For MA-enabled secure communications, the channel is reconfigurable by adjusting the positions of MAs. Denote the collections of the coordinates of $N$ MAs by ${\mathbf{T}}=[{\mathbf{t}}_1 \ldots {\mathbf{t}}_N]\in{\mathbbmss{R}}^{2\times N}$. Then, the transmitter-to-Bob and the transmitter-to-Eve channel vectors are given by ${\mathbf{h}}_{\rm{b}}(\mathbf{T})\in{\mathbbmss{C}}^{N\times1}$ and ${\mathbf{h}}_{\rm{e}}(\mathbf{T})\in{\mathbbmss{C}}^{N\times1}$, respectively, which are functions of $\mathbf{T}$. 

It is worth noting that the MA-based channel vectors are determined by the signal propagation environment and the positions of MAs. We consider the field-response based channel model given by ${\mathbf h}_{{i}}(\mathbf{T})=[h_{i}({\mathbf{t}}_1)\ldots h_{i}({\mathbf{t}}_N)]^{\mathsf{T}}$, where \cite{Ma2023}
{\setlength\abovedisplayskip{2pt}
\setlength\belowdisplayskip{2pt}
\begin{align}\label{Channel_Model}
h_{i}({\mathbf{x}})\triangleq\sum\nolimits_{l=1}^{L_i}\sigma_{l}^{i}{\rm{e}}^{{\rm{j}}\frac{2\pi}{\lambda}{\mathbf{x}}^{\mathsf{T}}{\bm\rho}_{l,i}},~i\in\{{\rm{b}},{\rm{e}}\},
\end{align}
}and where ${\bm\rho}_{l,i}=[\sin{\theta_{l,i}}\cos{\phi_{l,i}},\cos{\theta_{l,i}}]^{\mathsf{T}}$, $L_i$ is the number of resolvable paths, $\theta_{l,i}\in[0,\pi]$ and $\phi_{l,i}\in[0,\pi]$ are the elevation and azimuth angles of the $l$th path, respectively, $\sigma_{l}^{i}$ is the associated complex gain, and $\lambda$ is the wavelength. 

The transmitter employs the linear beamformer ${\mathbf{w}}\in{\mathbbmss{C}}^{N\times1}$ to construct the confidential signal ${\mathbf{s}}\in{\mathbbmss{C}}^{N\times1}$ from the data symbol $x\in{\mathbbmss{C}}$ which is considered to have zero mean and unit variance. The received signals at Bob and Eve are thus given by ${y}_i={\mathbf{h}}_i^{\mathsf{H}}(\mathbf{T}){\mathbf{s}}+n_i={\mathbf{h}}_i^{\mathsf{H}}(\mathbf{T}){\mathbf{w}}x+n_i$ for $i\in\{{\rm{b}},{\rm{e}}\}$, where $n_i\sim{\mathcal{CN}}(0,\sigma_i^2)$ is the thermal noise at node $i$ with $\sigma_i^2$ being the noise power. It is assumed that Eve is a registered user of the system but is distrusted by Bob. Therefore, the channel state information (CSI) of both Bob and Eve is available at the transmitter \cite{Chen2017}. To reveal the fundamental secrecy rate limit of the MA-enabled secure communication system, we further assume that the transmitter knows the CSI perfectly \cite{Zhu2023,Ma2022,Zhu2023_2,Cheng2023_1,Cheng2023_2,Sun2023,Ma2023}. The secrecy rate can be expressed as follows: \cite{Khisti2010}
{\setlength\abovedisplayskip{2pt}
\setlength\belowdisplayskip{2pt}
\begin{align}\label{Secrecy_Capacity}
{\mathcal{R}}_{\rm{s}}=\max\left\{\log_2\frac{1+\sigma_{\rm{b}}^{-2}\lvert{\mathbf{h}}_{\rm{b}}^{\mathsf{H}}(\mathbf{T}){\mathbf{w}}\rvert^2}
{1+\sigma_{\rm{e}}^{-2}\lvert{\mathbf{h}}_{\rm{e}}^{\mathsf{H}}(\mathbf{T}){\mathbf{w}}\rvert^2},0\right\}.
\end{align}
}Note that different from the conventional wiretap channel with FPAs, the secrecy rate for the MA-enabled wiretap channel shown in \eqref{Secrecy_Capacity} depends on the positions of MAs $\mathbf{T}$, which influence the channel vectors as well as the corresponding optimal beamformer.
\vspace{-10pt}
\subsection{Problem Formulation}
\vspace{-5pt}
In order to avoid the coupling effect between the antennas in the transmit region, a minimum distance $D$ is required between each pair of antennas, i.e., $\lVert {\mathbf{t}}_n-{\mathbf{t}}_{n'}\rVert\geq D$ for $n\ne n'$. Then, we aim to improve the secrecy performance of an MA-enabled wiretap channel by jointly optimizing the MA positions $\mathbf{T}$ and the transmit beamformer $\mathbf{w}$, subject to the minimum distance constraints on the MA positions. The joint design is performed under two criteria: 1) transmit power minimization; 2) secrecy rate maximization. We consider minimizing the transmit power with a preset secrecy rate target for the first design criterion, and the problem is formulated as
{\setlength\abovedisplayskip{2pt}
\setlength\belowdisplayskip{2pt}
\begin{align}\label{Power_Maximization_Problem}
\min\nolimits_{\mathbf{T},{\mathbf{w}}}\lVert\mathbf{w}\rVert^2~{\rm{s.t.}}~{\mathcal{R}}_{\rm{s}}\geq{\mathcal{R}},{\mathbf{t}}_n\in{\mathcal{C}},\lVert {\mathbf{t}}_n-{\mathbf{t}}_{n'}\rVert\geq D,
\tag{${\mathcal{P}}_1$}
\end{align}
}where ${\mathcal{R}}>0$ is the minimal secrecy rate requirement. We also wish to maximize the secrecy rate while guaranteeing the power budget. Accordingly, we formulate the problem as
{\setlength\abovedisplayskip{2pt}
\setlength\belowdisplayskip{2pt}
\begin{align}\label{Secrecy_Rate_Maximization_Problem}
\max\nolimits_{\mathbf{T},{\mathbf{w}}}{\mathcal{R}}_{\rm{s}}~{\rm{s.t.}}~\lVert\mathbf{w}\rVert^2\leq p,{\mathbf{t}}_n\in{\mathcal{C}},\lVert {\mathbf{t}}_n-{\mathbf{t}}_{n'}\rVert\geq D,
\tag{${\mathcal{P}}_2$}
\end{align}
}where $p$ is the power budget. 

Note that \eqref{Power_Maximization_Problem} and \eqref{Secrecy_Rate_Maximization_Problem} are non-convex optimization problems due to the non-convexity of $\mathcal{R}_{\rm{s}}$ with respect to $({\mathbf{T}},{\mathbf{w}})$ and the non-convex minimum distance constraint $\lVert {\mathbf{t}}_n-{\mathbf{t}}_{n'}\rVert\geq D$. Moreover, the beamformer $\mathbf{w}$ is coupled with $\mathbf{T}$, which makes \eqref{Power_Maximization_Problem} and \eqref{Secrecy_Rate_Maximization_Problem} challenging to solve.
\vspace{-10pt}
\section{Proposed Solution}
\vspace{-5pt}
\subsection{Transmit Power Minimization}
\vspace{-5pt}
\subsubsection{Transmit Beamforming Optimization}
\vspace{-5pt}
Due to the tight coupling of $\mathbf{w}$ and $\mathbf{T}$, we first explore the design of $\mathbf{w}$ by fixing $\mathbf{T}$. By introducing two auxiliary variables $\{{\mathbf v},\alpha\}$ subject to $\lVert{\mathbf v}\rVert^2=1$, $\lVert{\mathbf w}\rVert^2=\alpha$, and ${\mathbf w}=\sqrt{\alpha}{\mathbf v}$, the subproblem of optimizing $\mathbf w$ is given as follows:
{\setlength\abovedisplayskip{2pt}
\setlength\belowdisplayskip{2pt}
\begin{align}\label{Active_Beamforming_Opt_Power_Min_New}
\min_{{\mathbf v},\alpha}\alpha~~{\rm{s.t.}}~\frac{1\!+\!\alpha\sigma_{\rm{b}}^{-2}\lvert{\mathbf{h}}_{\rm{b}}^{\mathsf{H}}(\mathbf{T}){\mathbf{v}}\rvert^2}
{1\!+\!\alpha\sigma_{\rm{e}}^{-2}\lvert{\mathbf{h}}_{\rm{e}}^{\mathsf{H}}(\mathbf{T}){\mathbf{v}}\rvert^2}\!\geq\!2^{{\mathcal R}}, \lVert{\mathbf v}\rVert^2\!=\!1,\alpha\!>\!0.
\end{align}
}The first constraint in \eqref{Active_Beamforming_Opt_Power_Min_New} can be rewritten as
{\setlength\abovedisplayskip{2pt}
\setlength\belowdisplayskip{2pt}
\begin{align}\label{P5_Cons1_R1}
({{\mathbf v}^{\mathsf H}((1-2^{{\mathcal R}}){\mathbf I}+\alpha{\bm\Theta}){\mathbf v}})/
({{\mathbf v}^{\mathsf H}({\mathbf I}+\alpha\hat{\mathbf h}_{\rm e}\hat{\mathbf h}_{\rm e}^{\mathsf H}){\mathbf v}})\geq0.
\end{align}
}where $\hat{\mathbf{h}}_{i}\triangleq{\mathbf{h}}_{i}^{\mathsf{H}}(\mathbf{T})\sigma_i^{-1}$ for $i\in\{{\rm{b}},{\rm{e}}\}$ and ${\bm\Theta}\triangleq\hat{\mathbf h}_{\rm b}\hat{\mathbf h}_{\rm b}^{\mathsf H}-2^{{\mathcal R}}\hat{\mathbf h}_{\rm e}\hat{\mathbf h}_{\rm e}^{\mathsf H}\in{\mathbbmss C}^{N\times N}$. Since ${\mathbf v}\neq{\mathbf 0}$, we have ${\mathbf v}^{\mathsf H}({\mathbf I}+\alpha\hat{\mathbf h}_{\rm e}\hat{\mathbf h}_{\rm e}^{\mathsf H}){\mathbf v}>0$. Therefore, \eqref{P5_Cons1_R1} is equivalent to ${\mathbf v}^{\mathsf H}((1-2^{{\mathcal R}}){\mathbf I}+\alpha{\bm\Theta}){\mathbf v}\geq 0$. Let $\lambda_{\bm\Theta}$ denote the principal eigenvalue of ${\bm\Theta}$, and we have
{\setlength\abovedisplayskip{2pt}
\setlength\belowdisplayskip{2pt}
\begin{equation}\label{Power_Min_Eigenvalue}
\begin{split}
{\mathbf v}^{\mathsf H}((1-2^{{\mathcal R}}){\mathbf I}+\alpha{\bm\Theta}){\mathbf v}\leq 1-2^{{\mathcal R}}+\alpha\lambda_{\bm\Theta},
\end{split}
\end{equation}
}where the equality in \eqref{Power_Min_Eigenvalue} holds when $\mathbf v$ is the normalized principal eigenvector of ${\bm\Theta}$. Based on \eqref{Power_Min_Eigenvalue}, we note that $\lambda_{\bm\Theta}$ should satisfy $\alpha\lambda_{\bm\Theta}\geq2^{{\mathcal R}}-1>0$ when problem \eqref{Active_Beamforming_Opt_Power_Min_New} is feasible. This fact implies that $\lambda_{\bm\Theta}>0$ and $\alpha\geq\frac{2^{{\mathcal R}}-1}{\lambda_{\bm\Theta}}>0$, which means that the minimum value of $\alpha$ is given by $\alpha^{\star}=\frac{2^{{\mathcal R}}-1}{\lambda_{\bm\Theta}}$. Besides, when $\alpha=\alpha^{\star}$, the corresponding $\mathbf v$ is the normalized principal eigenvector of ${\bm\Theta}$, which is denoted by ${\mathbf v}^{\star}$. 

\begin{algorithm}[!t]
  \algsetup{linenosize=\tiny} \scriptsize
  \caption{Gradient-Based Algorithm for Solving \eqref{Power_Maximization_Problem}}
  \label{Algorithm1}
  \begin{algorithmic}[1]
    \STATE Initialize ${\mathbf{T}}^{0}=[{\mathbf{t}}_1^0 \ldots {\mathbf{t}}_N^0]$, the maximum iteration number $I$, step size $u_{\rm{ini}}$, the minimum tolerance step size $u_{\min}$, and set the current iteration $a=0$;
    \REPEAT
    \FORALL{$n=1:N$} 
    \STATE Compute the gradient value $\nabla_{{\mathbf{t}}_n^{a}}{f_n}$ and set $u=u_{\rm{ini}}$;
      \REPEAT
      \STATE Compute $\hat{\mathbf{t}}_n={\mathbf{t}}_n^{a}+u\cdot\nabla_{{\mathbf{t}}_n^{a}}{f_n}$ and set $u=u/2$;     
      \UNTIL{$\hat{\mathbf{t}}_n\in{\mathcal{S}}_n \& f_n(\hat{\mathbf{t}}_n)>f_n({\mathbf{t}}_n^{a})$ or $u<u_{\min}$};
      \STATE Set ${\mathbf{t}}_n^{a}=\hat{\mathbf{t}}_n$ and update ${\mathbf{t}}_n^{a+1}=\hat{\mathbf{t}}_n$;
    \ENDFOR
      \STATE Update $a=a+1$;
    \UNTIL{convergence or the maximum iteration number $I$ is reached};
    \STATE Update $\mathbf{w}$ by ${\mathbf w}^{\star}=\sqrt{\alpha^{\star}}{\mathbf v}^{\star}$.
  \end{algorithmic}
\end{algorithm} 

In the sequel, we derive a closed-form expression for $\lambda_N$. The eigenvalues of ${\bm\Theta}=\hat{\mathbf h}_{\rm b}\hat{\mathbf h}_{\rm b}^{\mathsf H}-2^{{\mathcal R}}\hat{\mathbf h}_{\rm e}\hat{\mathbf h}_{\rm e}^{\mathsf H}$ can be obtained from the characteristic equation as follows:
{\setlength\abovedisplayskip{2pt}
\setlength\belowdisplayskip{2pt}
\begin{align}
\det(\lambda{\mathbf I}-{\bm\Theta})=\det(\lambda{\mathbf I}-\hat{\mathbf h}_{\rm b}\hat{\mathbf h}_{\rm b}^{\mathsf H}+2^{{\mathcal R}}\hat{\mathbf h}_{\rm e}\hat{\mathbf h}_{\rm e}^{\mathsf H})=0,
\end{align}
}which along with the matrix determinant lemma \cite{Horn2012}, yields
{\setlength\abovedisplayskip{2pt}
\setlength\belowdisplayskip{2pt}
\begin{align}\label{Ch_Poly2_1}
(1+2^{{\mathcal R}}\hat{\mathbf h}_{\rm e}^{\mathsf H}(\lambda{\mathbf I}-\hat{\mathbf h}_{\rm b}\hat{\mathbf h}_{\rm b}^{\mathsf H})^{-1}\hat{\mathbf h}_{\rm e})\det(\lambda{\mathbf I}-\hat{\mathbf h}_{\rm b}\hat{\mathbf h}_{\rm b}^{\mathsf H})=0.
\end{align}
}Applying the Woodbury formula \cite{Horn2012} to \eqref{Ch_Poly2_1} gives
{\setlength\abovedisplayskip{2pt}
\setlength\belowdisplayskip{2pt}
\begin{align}\label{Ch_Poly2_2}
	\bigg(\!1\!+\!\frac{2^{{\mathcal R}}}{\lambda}\lVert\hat{\mathbf h}_{\rm e}\rVert^2\!+\!\frac{2^{{\mathcal R}}\lvert\hat{\mathbf h}_{\rm b}^{\mathsf H}\hat{\mathbf h}_{\rm e}\rvert^2}{\lambda(\lambda-\lVert\hat{\mathbf h}_{\rm b}\rVert^2)}\!\bigg)\!\det(\lambda{\mathbf I}\!-\!\hat{\mathbf h}_{\rm b}\hat{\mathbf h}_{\rm b}^{\mathsf H})\!=\!0.
\end{align}
}Based on the Sylvester's determinant identity \cite{Horn2012}, we obtain
{\setlength\abovedisplayskip{2pt}
\setlength\belowdisplayskip{2pt}
\begin{align}\label{Ch_Poly2_3}
	\det(\lambda{\mathbf I}-\hat{\mathbf h}_{\rm b}\hat{\mathbf h}_{\rm b}^{\mathsf H})=\lambda^{L-1}(\lambda-\lVert\hat{\mathbf h}_{\rm b}\rVert^2).
\end{align}
}Substituting \eqref{Ch_Poly2_3} into \eqref{Ch_Poly2_2} gives
{\setlength\abovedisplayskip{2pt}
\setlength\belowdisplayskip{2pt}
\begin{align}\label{Ch_Poly2_4}
	&\lambda^{L-2}((\lambda\!+\!2^{{\mathcal R}}\lVert\hat{\mathbf h}_{\rm e}\rVert^2)(\lambda\!-\!\lVert\hat{\mathbf h}_{\rm b}\rVert^2)\!+\!2^{{\mathcal R}}\lvert\hat{\mathbf h}_{\rm b}^{\mathsf H}\hat{\mathbf h}_{\rm e}\rvert^2)=0.
\end{align}
}We thus obtain the $N$ roots of the characteristic equation $\det(\lambda{\mathbf I}-{\bm\Theta})=0$, i.e., $\lambda_2=\ldots=\lambda_{N-1}=0$,
{\setlength\abovedisplayskip{2pt}
\setlength\belowdisplayskip{2pt}
\begin{align}
\lambda_{N}=({\sqrt{w_{\circ}}-w})/2,\lambda_{1} = ({-\sqrt{w_{\circ}}-w})/{2},
\end{align}
}where $w=2^{{\mathcal R}}\lVert\hat{\mathbf h}_{\rm e}\rVert^2-\lVert\hat{\mathbf h}_{\rm b}\rVert^2$, $w_{\circ}=w^2\!+\!2^{2+{\mathcal R}}G$, and $G=\lVert\hat{\mathbf h}_{\rm{b}}\rVert^2\lVert\hat{\mathbf h}_{\rm{e}}\rVert^2-\lvert\hat{\mathbf h}_{\rm b}^{\mathsf H}\hat{\mathbf h}_{\rm e}\rvert^2$. It is clear that $\lambda_1\lambda_N\leq0$, which yields $\lambda_1\leq\ldots\leq\lambda_N$. Consequently, the principal eigenvalue of ${\bm\Theta}$ is $\lambda_N$, i.e., $\lambda_{\bm\Theta}=\lambda_N$. 
\vspace{-10pt}
\subsubsection{Antenna Position Optimization}
\vspace{-5pt}
The above arguments imply that given the antenna position matrix $\mathbf{T}$, the minimal required power is given by $\alpha^{\star}=\frac{2^{{\mathcal R}}-1}{\lambda_N}$ along with the optimal beamformer ${\mathbf w}^{\star}=\sqrt{\alpha^{\star}}{\mathbf v}^{\star}$. It is worth noting that $\lambda_N$ is a function of $\mathbf{T}$. Based on this, we can equivalently transform problem \eqref{Power_Maximization_Problem} as follows:
{\setlength\abovedisplayskip{2pt}
\setlength\belowdisplayskip{2pt}
\begin{align}\label{P1_Transform}
\max\nolimits_{\mathbf{T},{\mathbf{t}}_n\in{\mathcal{C}},\forall n,\lVert {\mathbf{t}}_n-{\mathbf{t}}_{n'}\rVert\geq D,\forall n\ne n'}f(\mathbf{T})\triangleq\sqrt{w_{\circ}}-w.
\end{align}
}The main challenges in solving problem \eqref{P1_Transform} lie in the intractability of $f(\mathbf{T})$ and the tight coupling of $\{{\mathbf{t}}_n\}_{n=1}^{N}$. To this end, we propose an alternating optimization algorithm for solving \eqref{P1_Transform}. Specifically, we first divide $\mathbf{T}$ into $N$ blocks $\{{\mathbf{t}}_n\}_{n=1}^{N}$. Then, we solve $N$ subproblems of \eqref{P1_Transform}, which respectively optimize one transmit MA position ${\mathbf{t}}_n$, with all the other variables being fixed. The developed alternating optimization algorithm can obtain a (at least) locally optimal solution for \eqref{P1_Transform} or \eqref{Power_Maximization_Problem} by iteratively solving the above $N$ subproblems in an alternate manner.

We next consider the optimization of ${\mathbf{t}}_n$ with given $\{{\mathbf{t}}_{n'}\}_{n'\ne n}$. This subproblem is given as follows:
{\setlength\abovedisplayskip{2pt}
\setlength\belowdisplayskip{2pt}
\begin{align}\label{P1_Transform_Sub}
\max_{{\mathbf{t}}_n\in{\mathcal{S}}_n}f_n({\mathbf{t}}_n)\triangleq\sqrt{w_n^2({\mathbf{t}}_n)+2^{2+{\mathcal R}}G_n({\mathbf{t}}_n)}-w_n({\mathbf{t}}_n),
\end{align}
}where ${\mathcal{S}}_n\triangleq\{{\mathbf{x}}|{\mathbf{x}}\in{\mathcal{C}},\lVert {\mathbf{x}}-{\mathbf{t}}_{n'}\rVert\geq D,\forall n\ne n'\}$, $G_n(\mathbf{x})=h_{n}^{\rm{b}}(\mathbf{x})h_{n}^{\rm{e}}(\mathbf{x})-h_{n}^{1,2}(\mathbf{x})$, $w_n(\mathbf{x})=2^{{\mathcal R}}h_{n}^{\rm{e}}(\mathbf{x})-h_{n}^{\rm{b}}(\mathbf{x})$, $h_{n}^{i}(\mathbf{x})={\sigma_i^{-2}}(\lvert h_{i}({\mathbf{x}})\rvert^2+\sum\nolimits_{n'\ne n}\lvert h_{i}({\mathbf{t}}_{n'})\rvert^2)$ for $i\in\{{\rm{b}},{\rm{e}}\}$, and $h_{n}^{1,2}(\mathbf{x})=\sigma_{\rm{b}}^{-2}\sigma_{\rm{e}}^{-2}\lvert h_{\rm{b}}^{\mathsf{H}}({\mathbf{x}})h_{\rm{e}}({\mathbf{x}})+\sum\nolimits_{n'\ne n}h_{\rm{b}}^{\mathsf{H}}({\mathbf{t}}_{n'})h_{\rm{e}}({\mathbf{t}}_{n'})\rvert^2$. Due to the intractability of $f_n(\cdot)$, stationary points of subproblem \eqref{P1_Transform_Sub} can be found capitalizing on the gradient descent method with backtracking line search \cite{Boyd2004}. To this end, the gradient values are calculated as follows: 
{\setlength\abovedisplayskip{2pt}
\setlength\belowdisplayskip{0pt}
\begin{align}\label{Der1_P1}
\nabla_{{\mathbf{t}}_n}h_{n}^{i}=\sum_{l=1}^{L_i}\!\sum_{l'\ne l}\!\frac{\lvert\sigma_{l}^{i}\sigma_{l'}^{i}\rvert}{-\sigma_{i}^2\frac{\lambda}{4\pi}}\sin\!\left(\!\frac{2\pi}{\lambda}
{\mathbf{t}}_n^{\mathsf{T}}{\bm\rho}_{l,l'}^{i}\!+\!\theta_{l,l'}^{i}\!\right)\!{\bm\rho}_{l,l'}^{i},
\end{align}
}{\setlength\abovedisplayskip{0pt}
\setlength\belowdisplayskip{2pt}
\begin{align}
&\nabla_{{\mathbf{t}}_n}h_{n}^{1,2}\!=\!\sum_{l=1}^{L_{\rm{b}}}\!\sum_{l'=1}^{L_{\rm{e}}}\!\frac{\Re\{A_n\}\lvert\sigma_{l}^{\rm{b}}\sigma_{l'}^{\rm{e}}\rvert}{-\sigma_{\rm{b}}^{2}\sigma_{\rm{e}}^{2}\frac{\lambda}{4\pi}}
\sin\!\left(\!\frac{2\pi}{\lambda}{\mathbf{t}}_n^{\mathsf{T}}{\bm\rho}_{l,l'}\!+\!\theta_{l,l'}\!\right)\!{\bm\rho}_{l,l'}\nonumber\\
&\quad+\sum_{l=1}^{L_{\rm{b}}}\sum_{l'=1}^{L_{\rm{e}}}\frac{\Im\{A_n\}\lvert\sigma_{l}^{\rm{b}}\sigma_{l'}^{\rm{e}}\rvert}{\sigma_{\rm{b}}^{2}\sigma_{\rm{e}}^{2}\frac{\lambda}{4\pi}}
\cos\left(\frac{2\pi}{\lambda}{\mathbf{t}}_n^{\mathsf{T}}{\bm\rho}_{l,l'}+\theta_{l,l'}\right){\bm\rho}_{l,l'}\nonumber\\
&\quad+\sigma_{\rm{b}}^{-2}\lVert h_{\rm{b}}({\mathbf{t}}_n)\rVert^2\nabla_{{\mathbf{t}}_n}h_{n}^{\rm{e}}+\sigma_{\rm{e}}^{-2}\lVert h_{\rm{e}}({\mathbf{t}}_n)\rVert^2\nabla_{{\mathbf{t}}_n}h_{n}^{\rm{b}}\label{Der2_P1},
\end{align}
}where $A_n=\sum_{n'\ne n}h_{\rm{b}}^{\mathsf{H}}({\mathbf{t}}_{n'})h_{\rm{e}}({\mathbf{t}}_{n'})$, ${\bm\rho}_{l,l'}^{i}={\bm\rho}_{l,i}-{\bm\rho}_{l',i}$, ${\bm\rho}_{l,l'}={\bm\rho}_{l,{\rm{b}}}-{\bm\rho}_{l',{\rm{e}}}$, $\theta_{l,l'}^{i}=\angle{\sigma}_{l}^{i}-\angle{\sigma}_{l'}^{i}$, and $\theta_{l,l'}=\angle{\sigma}_{l}^{\rm{b}}-\angle{\sigma}_{l'}^{\rm{e}}$. Based on \eqref{Der1_P1} and \eqref{Der2_P1}, $\nabla_{{\mathbf{t}}_n}f_n$ can be derived after some basic mathematical manipulations.

The overall algorithm for solving \eqref{Power_Maximization_Problem} is given in Algorithm \ref{Algorithm1}. Since the transmit power is lower bounded, the convergence is guaranteed. Regarding the complexity of Algorithm \ref{Algorithm1}, it scales with ${\mathcal{O}}(N^3+IN(L_{\rm{b}}+L_{\rm{e}})^2\log_2\frac{1}{u_{\min}})$.
\vspace{-10pt}
\subsection{Secrecy Rate Maximization}
\vspace{-5pt}
\subsubsection{Transmit Beamforming Optimization}
\vspace{-5pt}
Having solved problem \eqref{Power_Maximization_Problem}, we now move to the secrecy rate maximization problem characterized in \eqref{Secrecy_Rate_Maximization_Problem}. Following the methodology used in the previous part, we first derive the optimal beamforming vector for a given antenna position matrix $\mathbf{T}$. The subproblem of optimizing $\mathbf{w}$ is given by
{\setlength\abovedisplayskip{2pt}
\setlength\belowdisplayskip{2pt}
\begin{align}\label{P2_Subproblem1}
{\mathbf{w}}^{\star}=\argmax\nolimits_{\lVert\mathbf{w}\rVert^2\leq p}\frac{1+\sigma_{\rm{b}}^{-2}\lvert{\mathbf{h}}_{\rm{b}}^{\mathsf{H}}(\mathbf{T}){\mathbf{w}}\rvert^2}
{1+\sigma_{\rm{e}}^{-2}\lvert{\mathbf{h}}_{\rm{e}}^{\mathsf{H}}(\mathbf{T}){\mathbf{w}}\rvert^2}.
\end{align}
}Under the positive secrecy rate constraint $\sigma_{\rm{b}}^{-2}\lvert{\mathbf{h}}_{\rm{b}}^{\mathsf{H}}(\mathbf{T}){\mathbf{w}}\rvert^2>\sigma_{\rm{e}}^{-2}\lvert{\mathbf{h}}_{\rm{e}}^{\mathsf{H}}(\mathbf{T}){\mathbf{w}}\rvert^2$, it is easily proved that $\lVert\mathbf{w}^{\star}\rVert^2=p$. On this basis, we reformulate problem \eqref{P2_Subproblem1} as follows:
{\setlength\abovedisplayskip{2pt}
\setlength\belowdisplayskip{2pt}
\begin{align}\label{P2_Subproblem1_Trans}
{\mathbf{w}}^{\star}=\argmax\nolimits_{\lVert\mathbf{w}\rVert^2= p}\frac{{\mathbf{w}}^{\mathsf{H}}(\hat{\mathbf{h}}_{\rm{b}}\hat{\mathbf{h}}_{\rm{b}}^{\mathsf{H}}
-\hat{\mathbf{h}}_{\rm{e}}\hat{\mathbf{h}}_{\rm{e}}^{\mathsf{H}}){\mathbf{w}}}{{\mathbf{w}}^{\mathsf{H}}(p^{-1}{\mathbf{I}}+\hat{\mathbf{h}}_{\rm{e}}\hat{\mathbf{h}}_{\rm{e}}^{\mathsf{H}}){\mathbf{w}}}.
\end{align}
}Problem \eqref{P2_Subproblem1_Trans} is a Rayleigh quotient, and it follows that 
{\setlength\abovedisplayskip{2pt}
\setlength\belowdisplayskip{2pt}
\begin{align}
{\mathbf{w}}^{\star}=\sqrt{p}(p^{-1}{\mathbf{I}}+\hat{\mathbf{h}}_{\rm{e}}\hat{\mathbf{h}}_{\rm{e}}^{\mathsf{H}})^{-\frac{1}{2}}{\mathbf{p}}
\lVert p^{-1}{\mathbf{I}}+\hat{\mathbf{h}}_{\rm{e}}\hat{\mathbf{h}}_{\rm{e}}^{\mathsf{H}})^{-\frac{1}{2}}{\mathbf{p}}\rVert^{-1},\nonumber
\end{align}
}where ${\mathbf{p}}\in{\mathbbmss{C}}^{N\times1}$ is the principal eigenvector of the matrix
{\setlength\abovedisplayskip{2pt}
\setlength\belowdisplayskip{2pt}
\begin{align}
{\bm\Delta}=(p^{-1}{\mathbf{I}}\!+\!\hat{\mathbf{h}}_{\rm{e}}\hat{\mathbf{h}}_{\rm{e}}^{\mathsf{H}})^{-\frac{1}{2}}
(\hat{\mathbf{h}}_{\rm{b}}\hat{\mathbf{h}}_{\rm{b}}^{\mathsf{H}}
\!-\!\hat{\mathbf{h}}_{\rm{e}}\hat{\mathbf{h}}_{\rm{e}}^{\mathsf{H}})
(p^{-1}{\mathbf{I}}\!+\!\hat{\mathbf{h}}_{\rm{e}}\hat{\mathbf{h}}_{\rm{e}}^{\mathsf{H}})^{-\frac{1}{2}}.\nonumber
\end{align}
}Besides, the secrecy rate is given by ${\mathcal{R}}_{\rm{s}}=\log_2(1+\mu_{\bm\Delta})$ when ${\mathbf{w}}={\mathbf{w}}^{\star}$, where $\mu_{\bm\Delta}$ is the principal eigenvalue of $\bm\Delta$.

\begin{figure}[!t]
    \centering
    \subfigbottomskip=0pt
	\subfigcapskip=-5pt
\setlength{\abovecaptionskip}{0pt}
    \subfigure[]
    {
        \includegraphics[height=0.18\textwidth]{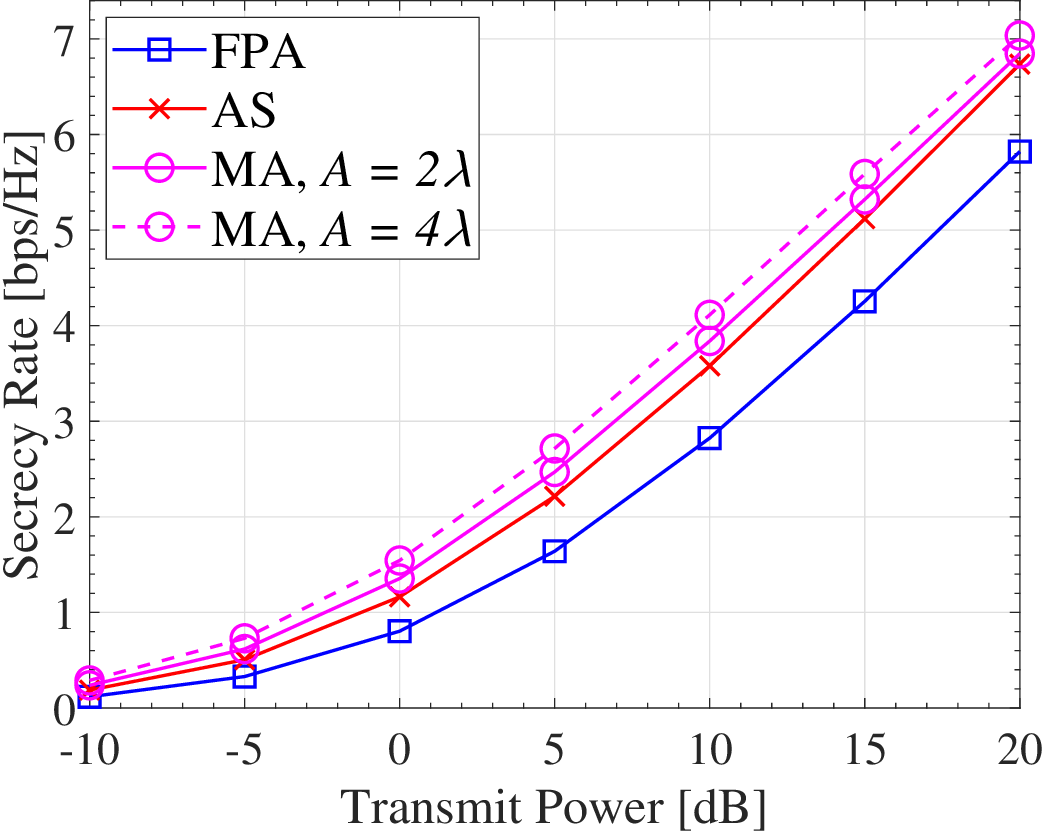}
	   \label{fig1a}	
    }
   \subfigure[]
    {
        \includegraphics[height=0.18\textwidth]{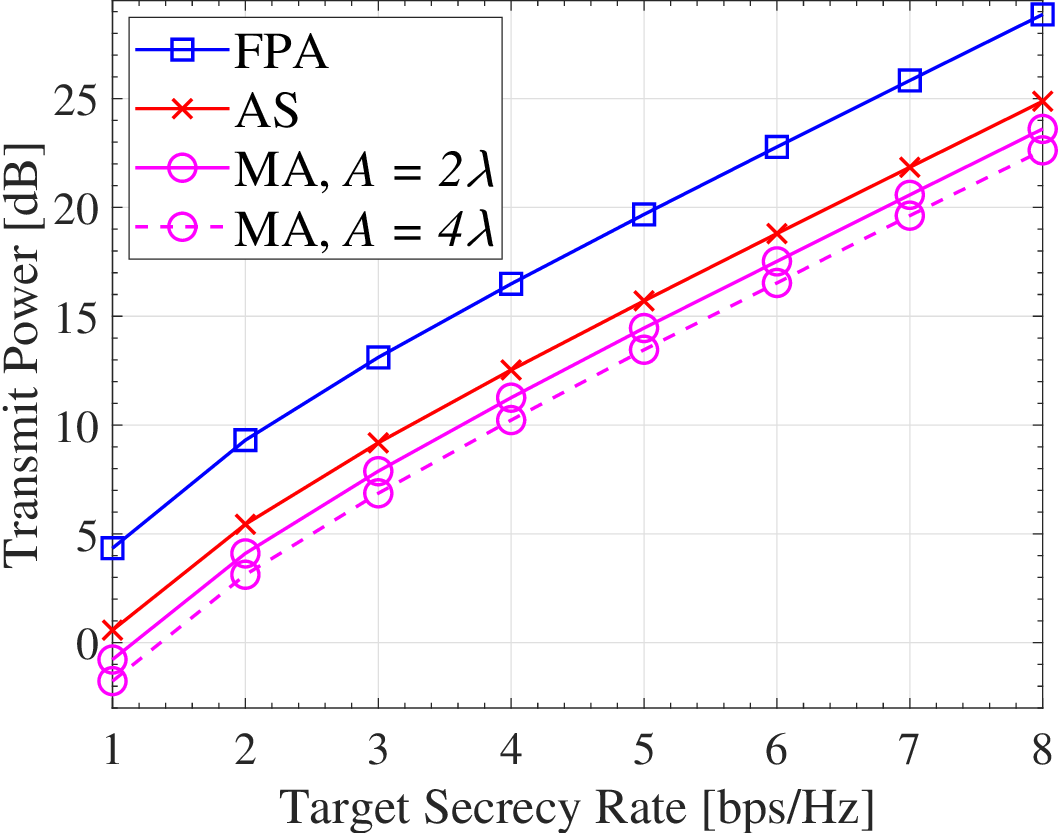}
	   \label{fig1b}	
    }
\caption{(a) Secrecy rate vs. the power budget and (b) Transmit power vs. the target secrecy rate.}
    \label{figure1}
    \vspace{-15pt}
\end{figure}

We next derive a closed-form expression for $\mu_{\bm\Delta}$. The eigenvalues of ${\bm\Delta}$ is obtained from the characteristic equation $\det({\bm\Delta}-\mu{\mathbf I})=0$, which yields $\det ( {\bm\Delta}_{\mu}-\hat{\mathbf{{h}}}_{\rm{b}}\hat{\mathbf{{h}}}_{\rm{b}}^{\mathsf H} ) =0$, where ${\bm\Delta}_{\mu}\triangleq{\mu}{p}^{-1}\mathbf{I}+(\mu+1)\hat{\mathbf{{h}}}_{\rm{e}}\hat{\mathbf{{h}}}_{\rm{e}}^{\mathsf H}$. Using the matrix determinant lemma and the Woodbury formula \cite{Horn2012}, we obtain
{\setlength\abovedisplayskip{2pt}
\setlength\belowdisplayskip{2pt}
\begin{align}\label{Ch_Poly_3}
\left(\!1\!-\!\frac{p}{\mu}\lVert \hat{\mathbf{{h}}}_{\rm{b}} \rVert ^2+\frac{p^2(1+\mu^{-1})\lvert \hat{\mathbf{{h}}}_{\rm{b}}^{\mathsf H}\hat{\mathbf{{h}}}_{\rm{e}} \rvert ^2}{\mu +p\left( \mu +1 \right) \lVert \hat{\mathbf{{h}}}_{\rm{e}} \rVert ^2}\!\right)\!
\det({\bm\Delta}_{\mu})=0.
\end{align}
}It follows from Sylvester's determinant identity \cite{Horn2012} that
{\setlength\abovedisplayskip{2pt}
\setlength\belowdisplayskip{2pt}
\begin{align}\label{Ch_Poly_4}
\det({\bm\Delta}_{\mu})=({\mu}{p}^{-1})^{N-2}{\mu}{p}^{-1}({\mu}{p}^{-1}\!+\!\left( \mu+1 \right) \lVert \hat{\mathbf{h}}_{\rm{e}} \rVert ^2 ).
\end{align}
}Substituting \eqref{Ch_Poly_4} into \eqref{Ch_Poly_3} and solving the resultant equation, we obtain the $N$ roots of the characteristic equation $\det({\bm\Delta}-\mu{\mathbf I})=0$, i.e., $\mu_1=\ldots=\mu_{N-2}=0$,
{\setlength\abovedisplayskip{2pt}
\setlength\belowdisplayskip{2pt}
\begin{align}
\mu_{N-1}=\frac{g-\sqrt{g^2+4q}}{2(1+p\lVert\hat{\mathbf h}_{\rm{e}}\rVert^2)},~\mu_N=\frac{g+\sqrt{g^2+4q}}{2(1+p\lVert\hat{\mathbf h}_{\rm{e}}\rVert^2)},
\end{align}
}where $G=\lVert\hat{\mathbf h}_{\rm{b}}\rVert^2\lVert\hat{\mathbf h}_{\rm{e}}\rVert^2-\lvert\hat{\mathbf h}_{\rm b}^{\mathsf H}\hat{\mathbf h}_{\rm e}\rvert^2$, $g=p\lVert\hat{\mathbf h}_{\rm{b}}\rVert^2-p\lVert\hat{\mathbf h}_{\rm{e}}\rVert^2+p^2G$, and $q=p^2(1+p\lVert\hat{\mathbf h}_{\rm{e}}\rVert^2)G$. It is clear that $\mu_{N-1}\mu_N=\frac{-2q}{1+p\lVert\hat{\mathbf h}_{\rm{e}}\rVert^2}\leq0$, which yields $\mu_1\leq\ldots\leq\mu_N$. Therefore, $\mu_N$ is the principal eigenvalue of $\bm\Delta$, i.e., $\mu_{\bm\Delta}=\mu_N$.
\vspace{-10pt}
\subsubsection{Antenna Position Optimization}
\vspace{-5pt}
The above arguments imply that given the antenna position matrix $\mathbf{T}$, the maximal secrecy rate is given by $\log_2(1+\mu_N)$. It is worth noting that $\mu_N$ is a function of $\mathbf{T}$. Based on this, we can equivalently transform problem \eqref{Secrecy_Rate_Maximization_Problem} as follows:
{\setlength\abovedisplayskip{2pt}
\setlength\belowdisplayskip{2pt}
\begin{align}\label{P2_Transform}
\max_{\mathbf{T},{\mathbf{t}}_n\in{\mathcal{C}},\forall n,\lVert {\mathbf{t}}_n-{\mathbf{t}}_{n'}\rVert\geq D,\forall n\ne n'}\tilde{f}(\mathbf{T})\triangleq\frac{g+\sqrt{g^2+4q}}{2(1+p\lVert\hat{\mathbf h}_{\rm{e}}\rVert^2)}.
\end{align}
}Note that a suboptimal solution of problem \eqref{P2_Transform} can be found by following similar steps as those outlined Algorithm \ref{Algorithm1}. Due to the page limitations, further discussions are skipped here and left as a potential direction for future work.

\vspace{-10pt}
\section{Numerical Results}
\vspace{-5pt}
In this section, numerical results are provided to validate the effectiveness of our proposed algorithms for improving the secrecy performance of the MA-enabled secure communication system. In the simulation, we consider a transmitter with $N = 4$ MAs. The transmit region are set as a square area with size $A\times A$. We consider the geometry channel model, where $L_{\rm{b}}=L_{\rm{e}}=4$, $\sigma_{l}^{i}\sim{\mathcal{CN}}(0,1/L_{i})$ for $l=1,\ldots,L_i$ and $i\in\{{\rm{b}},{\rm{e}}\}$, and the elevation and azimuth angles are randomly set within $[0,\pi]$. The minimum distance between MAs is set as $D = \frac{\lambda}{2}$. The noise power is $\sigma_i^2=1$ for $i\in\{{\rm{b}},{\rm{e}}\}$. We set $u_{\min}=10^{-3}$, $u_{\rm{ini}}=10$, and $I=20$. The following numerical results are averaged over 1000 independent channel realizations with randomly initialized optimization variables. 

Next, we compare the performance of our proposed algorithms with the following benchmark schemes: {\romannumeral1}) FPA: the transmitter is equipped with FPA-based uniform linear array (ULA) with $N$ antennas spaced by $\frac{\lambda}{2}$. {\romannumeral2}) AS: The transmitter is equipped with FPA-based ULA with $2N$ antennas spaced by $\frac{\lambda}{2}$, where $N$ transmit antennas are selected via exhaustive search to maximize $f(\mathbf{T})$ or $\tilde{f}(\mathbf{T})$.

In {\figurename} {\ref{fig1a}}, we show the secrecy rate of the proposed and benchmark schemes versus the transmit power $p$. It is observed that with the same power, our proposed algorithm can achieve a larger secrecy rate as compared to the schemes with FPAs. For the case with $p=5$ dB, the proposed scheme with $A=4\lambda$ has 65.7\% and 22.6\% performance improvements over the FPA and AS schemes, respectively. Besides, the rate gap increases with the area size of the transmit region $\mathcal{C}$. {\figurename} {\ref{fig1b}} illustrates the average transmit power versus the target secrecy rate value $\mathcal{R}$. It is observed that as the minimum required rate value increases, the system consumes more transmit power to satisfy the more rigorous secure requirements. 
We also observe that the proposed scheme outperforms the baseline schemes for the entire range of $\mathcal{R}$.

\begin{figure}[!t]
    \centering
    \subfigbottomskip=0pt
	\subfigcapskip=-5pt
\setlength{\abovecaptionskip}{0pt}
    \subfigure[$p=10$ dB.]
    {
        \includegraphics[height=0.175\textwidth]{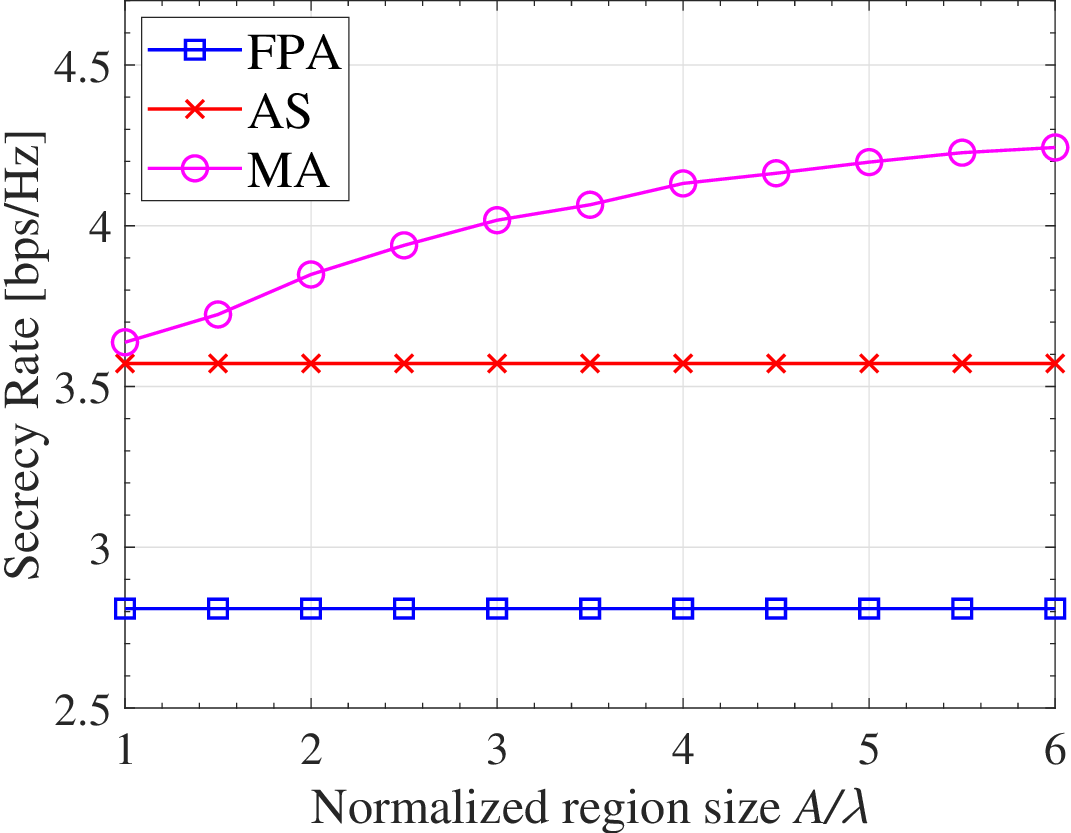}
	   \label{fig2a}	
    }
   \subfigure[${\mathcal{R}}=4$ bps/Hz.]
    {
        \includegraphics[height=0.175\textwidth]{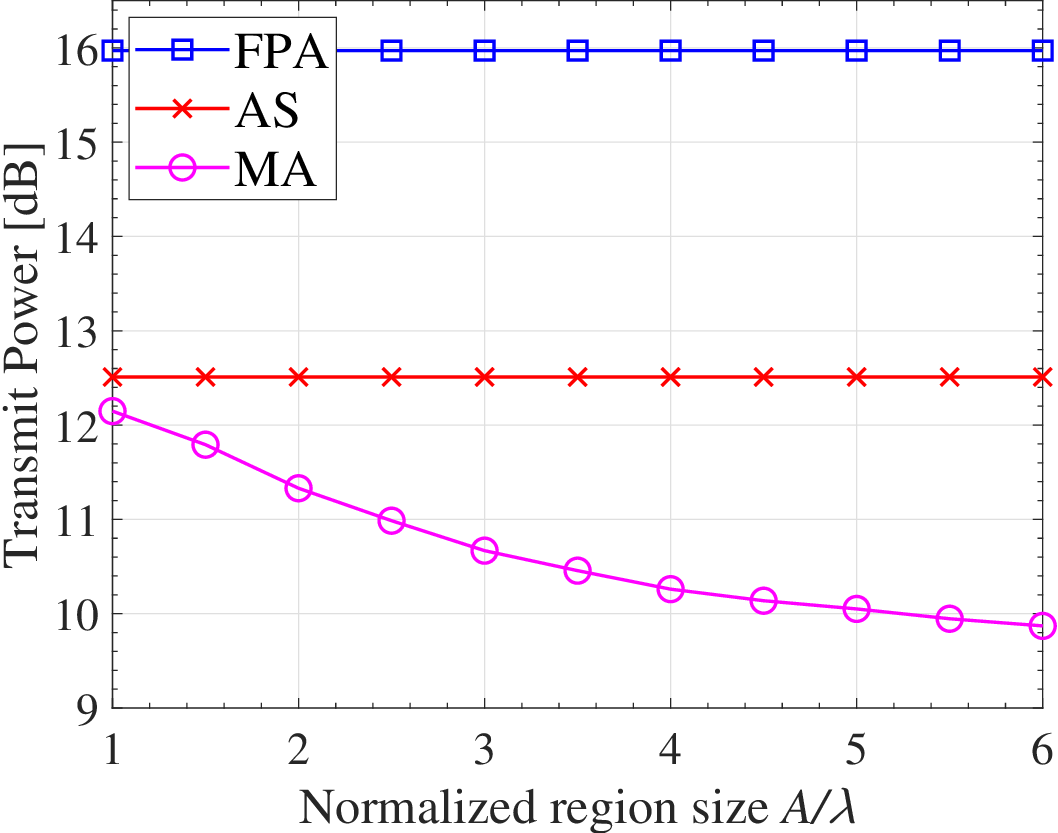}
	   \label{fig2b}	
    }
\caption{Secrecy performance vs. normalized region size.}
    \label{figure2}
    \vspace{-15pt}
\end{figure}
In {\figurename} {\ref{fig2a}} and {\figurename} {\ref{fig2b}}, we show the secrecy rate and the required transmit power versus the normalized region size $A/\lambda$, respectively, for the proposed MA-enabled wiretap system and the benchmark schemes. For both graphs, it is observed that the proposed schemes with MAs outperform FPA systems in terms of secrecy rate or required power. Furthermore, the performance gain increases with the region size, but its trend gradually steps down, from which we can infer that the best secrecy performance of MA-enabled communication systems can be achieved with a finite transmit region. The above results indicate that our proposed algorithms are able to reshape the wiretap channel into a more favorable condition for secrecy rate maximization.
\vspace{-10pt}
\section{Conclusion}
\vspace{-5pt}
In this paper, we proposed an MA-enabled secure transmission system to improve secrecy performance by exploiting the antenna position optimization. We investigated the joint optimization of transmit MA positions and transmit beamformer under transmit power minimization and secrecy rate maximization criteria. We proposed an efficient alternating optimization algorithm for each design problem. Numerical results revealed that the proposed MA-based architecture provides more DoFs for improving the secrecy rate and outperforms conventional FPA-based ones.

\end{document}